\documentclass[acus]{JAC2003}

\usepackage{bm}
\usepackage{multicol}
\usepackage{graphicx}
\usepackage{latexsym}
\usepackage{amsmath}
\usepackage[draft]{hyperref}

\hypersetup{linktocpage=false,
            pdfborder={0 0 0},
            colorlinks=false,
            unicode=true}

\usepackage{txfonts}
\usepackage{cancel}
\usepackage{bbding}
\usepackage{subfig}
\usepackage{float}
\usepackage{subfloat}
\usepackage{caption}
\usepackage[section]{placeins}
\DeclareGraphicsExtensions{.pdf,.png,.jpg}

\usepackage[hang]{footmisc}
\begin{document}

\title{LONG TERM EVOLUTION OF PLASMA WAKEFIELDS%
		\thanks{Work supported by the National Science Foundation under NSF-PHY-0936278 and the US Department of Energy under DE-SC0010012}\\[-.8\baselineskip]}

\author{A.A. Sahai\thanks{aakash.sahai@duke.edu}, T.C. Katsouleas,\\
		Electrical Engineering, Duke University, Durham, NC, 27708 USA\\
		Frank S. Tsung, Warren B. Mori\\
		Physics and Astronomy, UCLA, Los Angeles, CA, 90095 USA}

\maketitle

\begin{abstract}
We study the long-term evolution (LTE) of plasma wakefields over multiple plasma-electron periods and few plasma-ion periods, much less than a recombination time. The evolution and relaxation of such a wakefield-perturbed plasma over these timescales has important implications for the upper limits of repetition-rates in plasma colliders. Intense fields in relativistic lasers (or intense beams) create plasma wakefields (modes around $\omega_{pe}$) by transferring energy to the plasma electrons. Charged-particle beams in the right phase may be accelerated with acceleration/focusing gradients of tens of GeV/m. However, wakefields leave behind a plasma not in equilibrium, with a relaxation time of multiple plasma-electron periods. Ion motion over ion timescales, caused by energy transfer from the driven plasma-electrons to the plasma-ions can create interesting plasma states. Eventually during LTE, the dynamics of plasma de-coheres (multiple modes through instability driven mixing), thermalizing into random motion (second law of thermodynamics), dissipating energy away from the wakefields. Wakefield-drivers interacting with such a relativistically hot-plasma lead to plasma wakefields that differ from the wakefields in a cold-plasma.
\end{abstract}

\section{Introduction}
Intense laser and particle beams interacting with plasma can force plasma electrons to relativistic energies. Under the appropriate plasma conditions, energy gained by the plasma electrons can be used to create short-lived coherent structures in the plasma. These structures (such as linear electron waves and bubble) have been extensively studied and employed as accelerating and focussing systems for charged-particle beams. The structures created in the plasma have spatio-temporal scales that are defined by the characteristic of the plasma. The spatial scales by the plasma skin-depth, $\frac{c}{\omega_{pe}}$ ($\omega_{pe}=\sqrt{4\pi e^2/m_e}\sqrt{n_e/\gamma_e}$). And the temporal scales by the plasma frequency, $\frac{1}{\omega_{pe}}$. It should be noted that the wakefield theory assumes fixed background ions. In addition to this, the plasma structures are used only over a few plasma electron periods.  However, when these plasma structures evolve over time, ion-dynamics which occurs over ion-acoustic wave times $\frac{1}{\omega_{ia}}$ ($\omega_{ia} = \omega_{pe} \sqrt{m_e/M_i}$) becomes significant. For instance, in electron-proton plasma, $\omega_{pe} = 43~\omega_{ia}$. If plasma density, $n_e=10^{18}$, $\frac{1}{\omega_{pe}}\simeq 20fs$ and  $\frac{1}{\omega_{ia}}\simeq 1ps$. These timescales are important to understand the upper limits of the repetition-rate (and hence luminosity) in a future plasma collider \cite{LTE-2012}. Also, highly relativistic plasmas are created in astrophysical entities such as Active Galactic Nuclei (AGN such as Centaurus-A (NGC-5128), M-87) and lab-based hot plasmas can better explain phenomenon such as the relativistic jets emanating from AGNs. These particle jets interact with surrounding gravitationally heated hot plasma and can exhibit phenomenon such as self-focussing, hosing and radiation generation.

The plasma wave is {\it coherent} under the cold-plasma approximation as seen from its dispersion characteristic, $\omega^2=\omega^2_{pe}+3k^2v_{th}^2$ which under, $v_{phase}\gg v_{th}$ is excited over a narrow-band mode centered around $\omega_{pe}$. However, if the plasma is thermally inhomogeneous or the electrons have a large thermal spread the plasma wave mode begins to lose its narrowband characteristic. When a plasma wave evolves in time, plasma spatial inhomogeneities, driver spatial non-uniformity and ion motion introduce a new set of frequencies into the plasma. With non-linearity and instability driven phase mixing between these modes, the plasma wave further de-coheres and equilibriates into random thermal motion. This increase in the entropy of the system follows from the second law of thermodynamics which makes the energy distribution equipartitioned between all possible modes of a system, driving it towards equilibrium.

In a future plasma collider, the need for high luminosity would require each plasma stage to be re-used within a finite interval (depending upon the drive bunch-train design).  The fraction of drive-bunch energy coupled into the plasma that is absorbed by the accelerated-bunch has been shown to be $\simeq 50\%$ under optimal non-linear beam-loading which implies significant energy leftover in the accelerating plasma structure. Hence the next drive bunch would be affected by the thermalization of the wake from the previous pulses. In this paper we discuss the long-term evolution of a wake-perturbed  plasma in terms of its temperature and density profile. We look at the electron momenta to determine if unabsorbed energy from a previous drive-bunch could heat the plasma to relativistic temperatures ($k_BT_e\simeq 0.5 MeV$). We use simulations to estimate the plasma diffusion timescales. Past experimental work have imaged about 10 plasma periods behind the driver \cite{LTE-imaging}.

%%%%%%%%%%%%%%%%%%%%%%%%%%%
% ENERGY in a plasma wave
%%%%%%%%%%%%%%%%%%%%%%%%%%%
\section{Energy in a plasma wave}

The planar electron plasma wave in 1-D can be denoted as $\vec{E}_{pw}(x)=\hat{x} ~ E^0_{pw}  ~ e^{{\bf i}(\omega_{pe}t-k_{pe}x)}$. The force equation of the plasma wave acting on a single plasma electron is $m_e\frac{d\vec{v_e}}{dt}=-e\vec{E}_{pw}$. Therefore, $\vec{v_e}=\frac{-eE^0_{pw}}{m_e}\int e^{{\bf i}(\omega_{pe}t-k_{pe}x)}\hat{x}=\frac{e\vec{E}_{pw}}{m_e\omega_{pe}} e^{ {\bf i} \pi/2 }$ and $|\vec{v_e}|/c=\frac{e|\vec{E}_{pw}|}{m_ec\omega_{pe}}$. From the electron velocity we can estimate the kinetic energy of the single electron in the plasma wave, $\mathcal{E}_k = \frac{1}{2} ~ \left(\frac{e|\vec{E}_{pw}|}{m_ec\omega_{pe}}\right)^2 ~ m_e c^2$. If all the plasma electrons are picked up by the plasma wave, we can estimate the kinetic energy density of the electrons in the plasma wave as $\mathcal{E}_k/unit ~ volume = \frac{1}{2} ~ \left(\frac{e|\vec{E}_{pw}|}{m_ec\omega_{pe}}\right)^2 ~ m_e c^2 ~ n_e$. The field energy density in the plasma wave fields is $\mathcal{W}_f = \frac{|\vec{E}_{pw}|^2}{8\pi}=\left(\frac{e|\vec{E}_{pw}|}{m_ec\omega_{pe}}\right)^2~\frac{m_e^2c^2\omega_{pe}^2}{8\pi e^2}$. Using $\omega_{pe}^2=\frac{4\pi n_e e^2}{m_e}$, we have $\mathcal{W}_f = \frac{1}{2} ~ \frac{m_e\omega_{pe}^2}{4\pi n_e e^2} ~\left(\frac{e|\vec{E}_{pw}|}{m_ec\omega_{pe}}\right)^2 ~ m_ec^2n_e = \frac{1}{2} ~ \left(\frac{e|\vec{E}_{pw}|}{m_ec\omega_{pe}}\right)^2 ~ m_ec^2n_e$. As $\mathcal{E}_k=\mathcal{W}_f$, there is an energy exchange between the kinetic energy of the oscillating electrons and the electric field of the wave.

If the beam-loading of the plasma wave by an accelerated beam is $50\%$, the energy leftover in the plasma can be estimated as $\mathcal{E}_{plasma}/unit ~ volume = \frac{1}{2} ~ \left(\frac{e|\vec{E}_{pw}|}{m_ec\omega_{pe}}\right)^2 ~ m_e c^2 ~ n_e$. We study the density and temperature distribution of the plasma (electrons and ions) over $\simeq 10000$ cycle of $\frac{1}{\omega_{pe}}$ and $\simeq 1000$ cycle of $\frac{1}{\omega_{ia}}$.

%%%%%%%%%%%%%%%%%%%%%%%%%%%
% LTE SIMULATIONS
%%%%%%%%%%%%%%%%%%%%%%%%%%%
\section{Simulations of long term evolution of laser driven wakefield}

To study the LTE of a laser wakefield in the plasma, we use $2\frac{1}{2}D$ OSIRIS\cite{osiris} PIC code with Eulerian specification of the plasma dynamics. We initialize the plasma density to $0.01n_c$ with background ions of $10m_p$ and pre-ionized singly-charged state. We have intentionally chosen a relatively high density plasma because $\omega_{pe}=\omega_0\times\sqrt{n_e/n_c}$. We resolve and reference the real time in simulation to the laser period $2\pi/\omega_0$ thereby the dynamics within a single plasma cycle is simulated in just $\sqrt{n_c/n_e}=10$ laser cycles.  We discretize the space with 20 cells per skin-depth ($c/\omega_{pe}$) in the longitudinal and 10 cells per $c/\omega_{pe}$ in the transverse direction. The simulation space size is chosen as 300 $c/\omega_{pe}$ x 600 $c/\omega_{pe}$. We use absorbing boundary conditions for fields and particles of all species. The laser pulse is chosen to be a circularly polarized Gaussian pulse with normalized vector potential $a_0=4.0$ with pulse width of $10fs$.

\begin{figure}[ht]
	\centering
   	\includegraphics[width=\columnwidth]{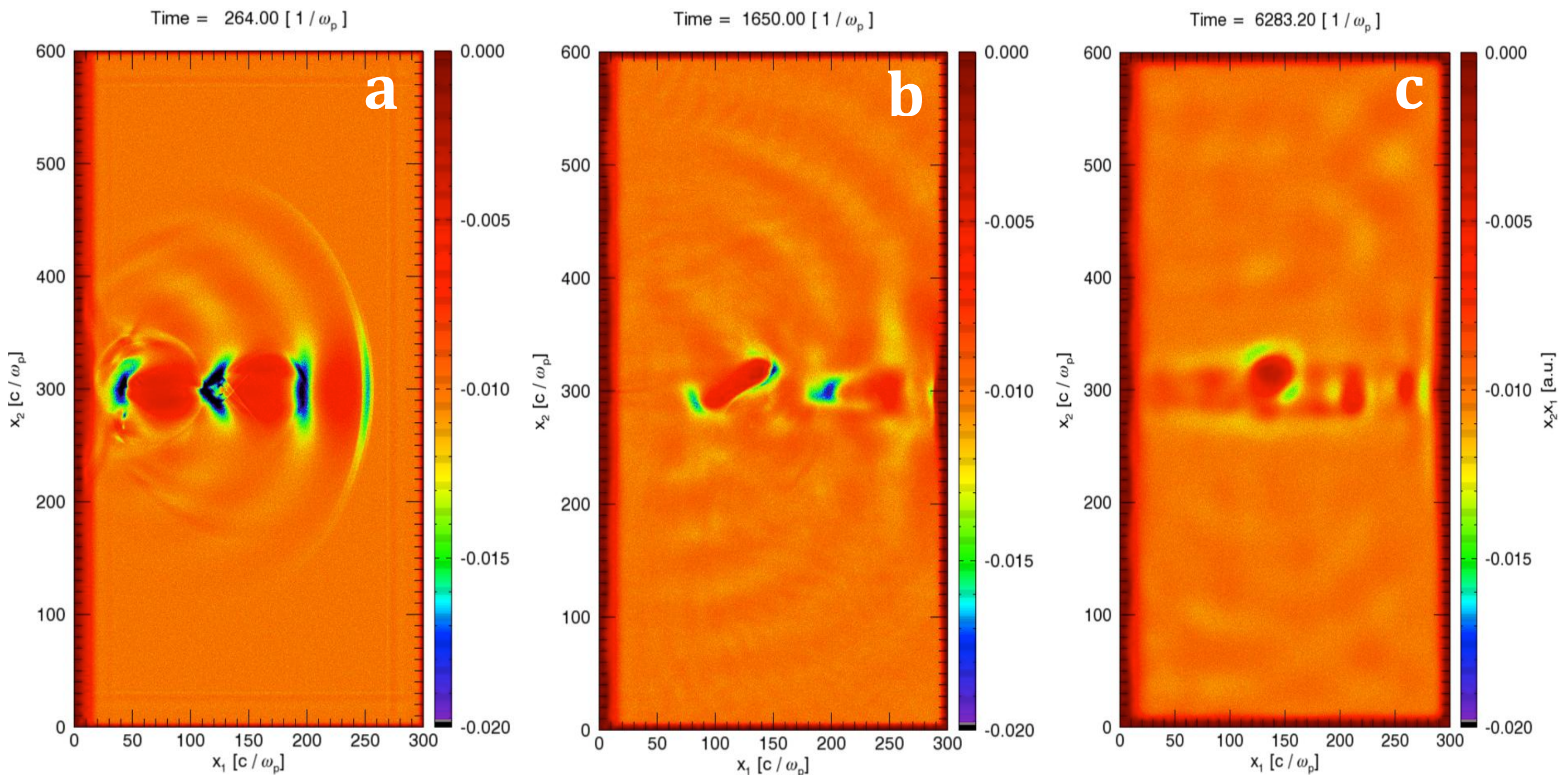}
	\caption{Laser wakefield LTE snapshots of {\bf plasma electron density in real space} at times, t = {\bf (a)} 264$\frac{1}{\omega_{pe}}$ {\bf (b)} 1650$\frac{1}{\omega_{pe}}$ {\bf (c)} 6283.20$\frac{1}{\omega_{pe}}$.}
	\label{LTE-electron-density-evolution}
\end{figure}

\begin{figure}[ht]
	\centering
   	\includegraphics[width=\columnwidth]{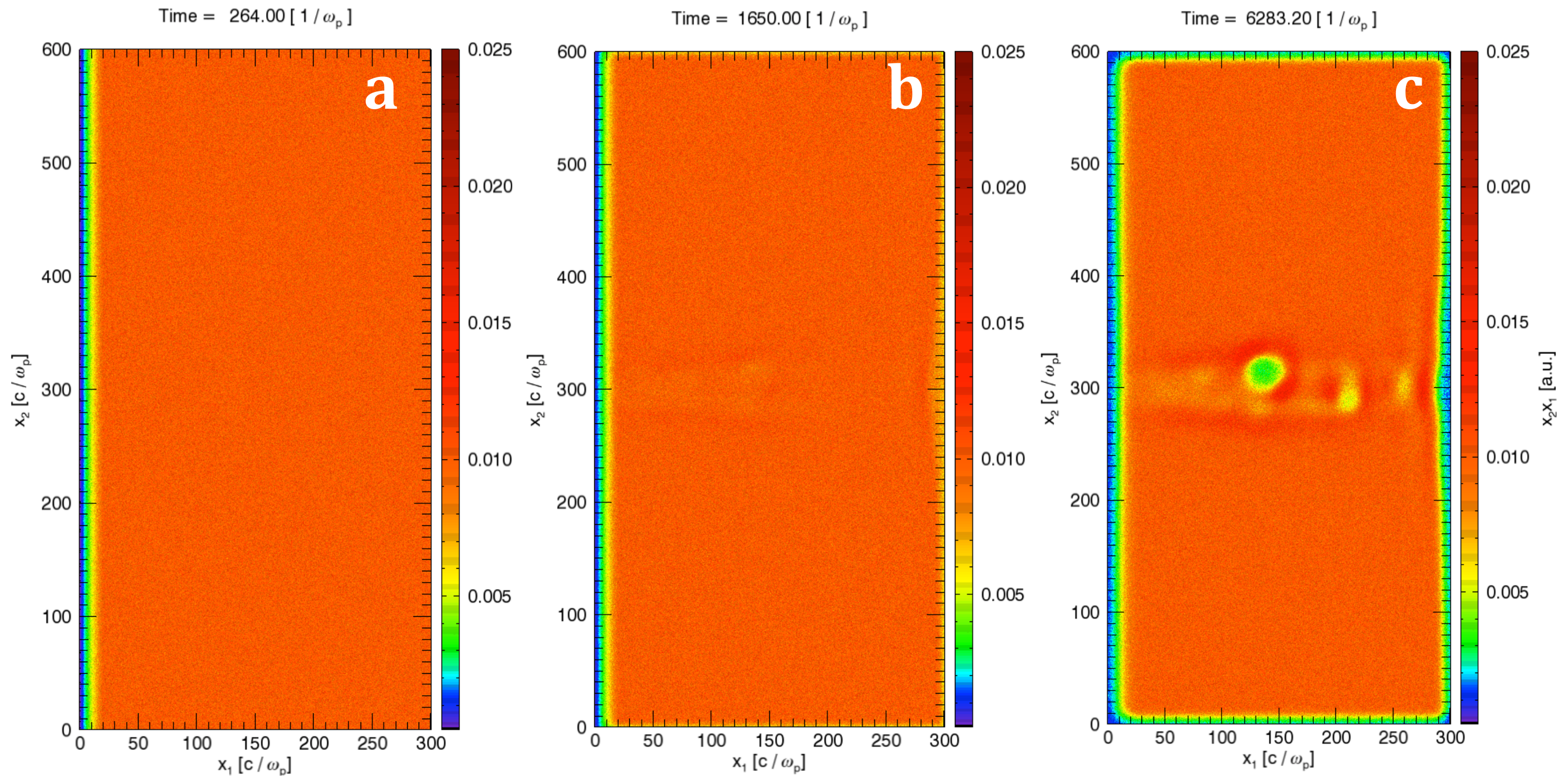}
	\caption{LTE snapshots of {\bf plasma ion density in real space} at times, t = {\bf (a)} 264$\frac{1}{\omega_{pe}}$ {\bf (b)} 1650$\frac{1}{\omega_{pe}}$ {\bf (c)} 6283.20$\frac{1}{\omega_{pe}}$. We can observe onset of ion motion in {\bf (b)}.}
	\label{LTE-ion-density-evolution}
\end{figure}

From the simulations we observe that a plasma bubble is excited by the laser pulse, as seen in Fig.~\ref{LTE-electron-density-evolution}[a] at $264\frac{1}{\omega_{pe}}$. The ions are stationary as seen from the ion density in Fig.~\ref{LTE-ion-density-evolution}[a]. We observe that the bubble oscillates for about 20 plasma periods. The more the oscillations are delayed in time, the more the bubble's transverse size decreases and the electron sheath around the bubble curves with smaller radii towards the back of the bubble. We also observe at the end of the bubble oscillations that the density of electrons in the sheath around the bubble reduces and the electron density is modulated further away from the axis of the bubble in the transverse direction. Such a weak bubble with curved electron sheath may be observed around $200\frac{c}{\omega_{pe}}$ in Fig.~\ref{LTE-electron-density-evolution}[b].

\begin{figure}
	\centering
   	\includegraphics[width=\columnwidth]{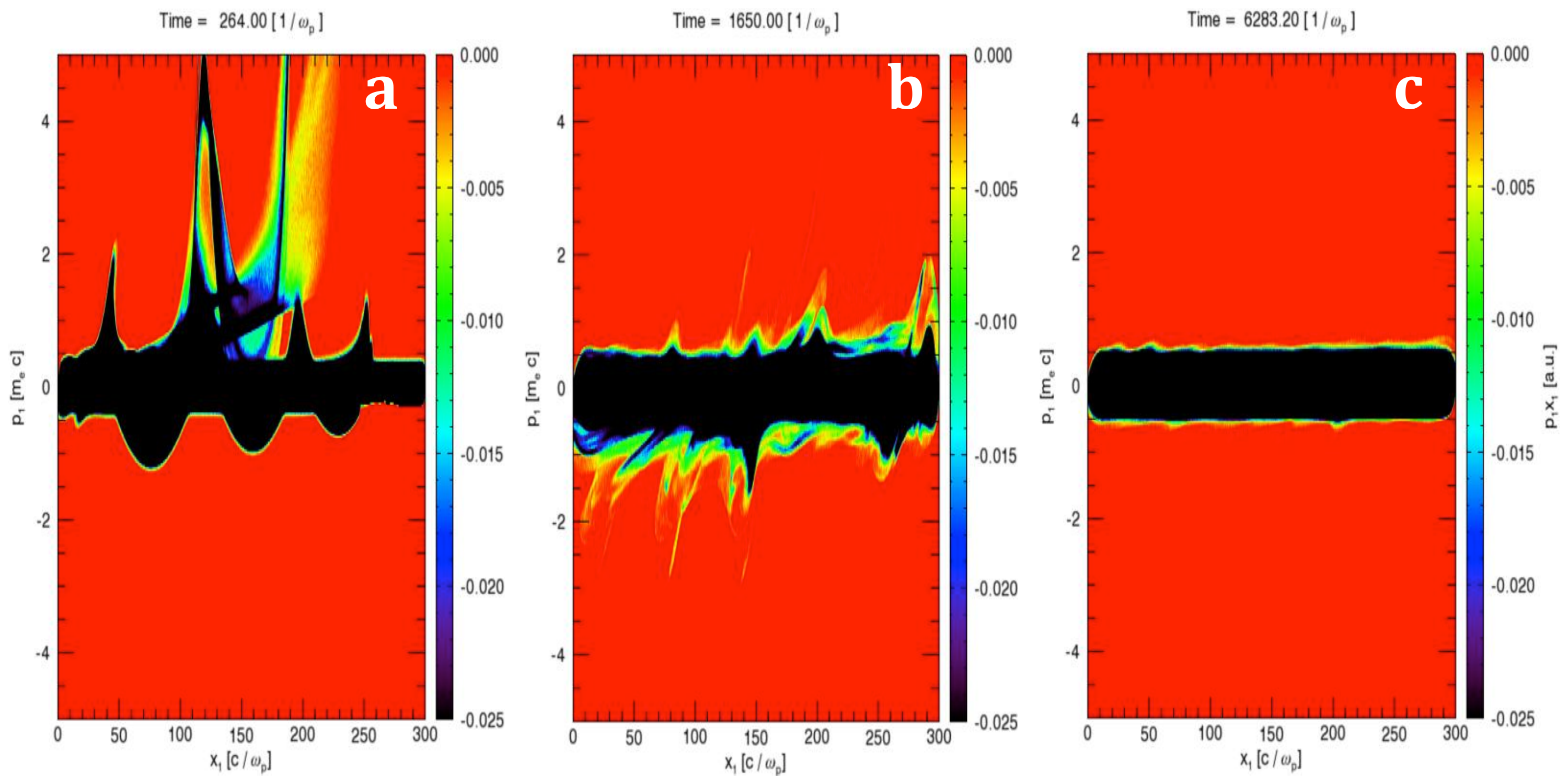}
	\caption{LTE snapshots of {\bf electron longitudinal momentum phase space} along longitudinal dimension at times, t= {\bf (a)} 264$\frac{1}{\omega_{pe}}$, {\bf (b)} 1650$\frac{1}{\omega_{pe}}$, {\bf (c)} 6283.20$\frac{1}{\omega_{pe}}$.}
	\label{LTE-longitudinal-momentum-evolution}
\end{figure}

\begin{figure}
	\centering
   	\includegraphics[width=\columnwidth]{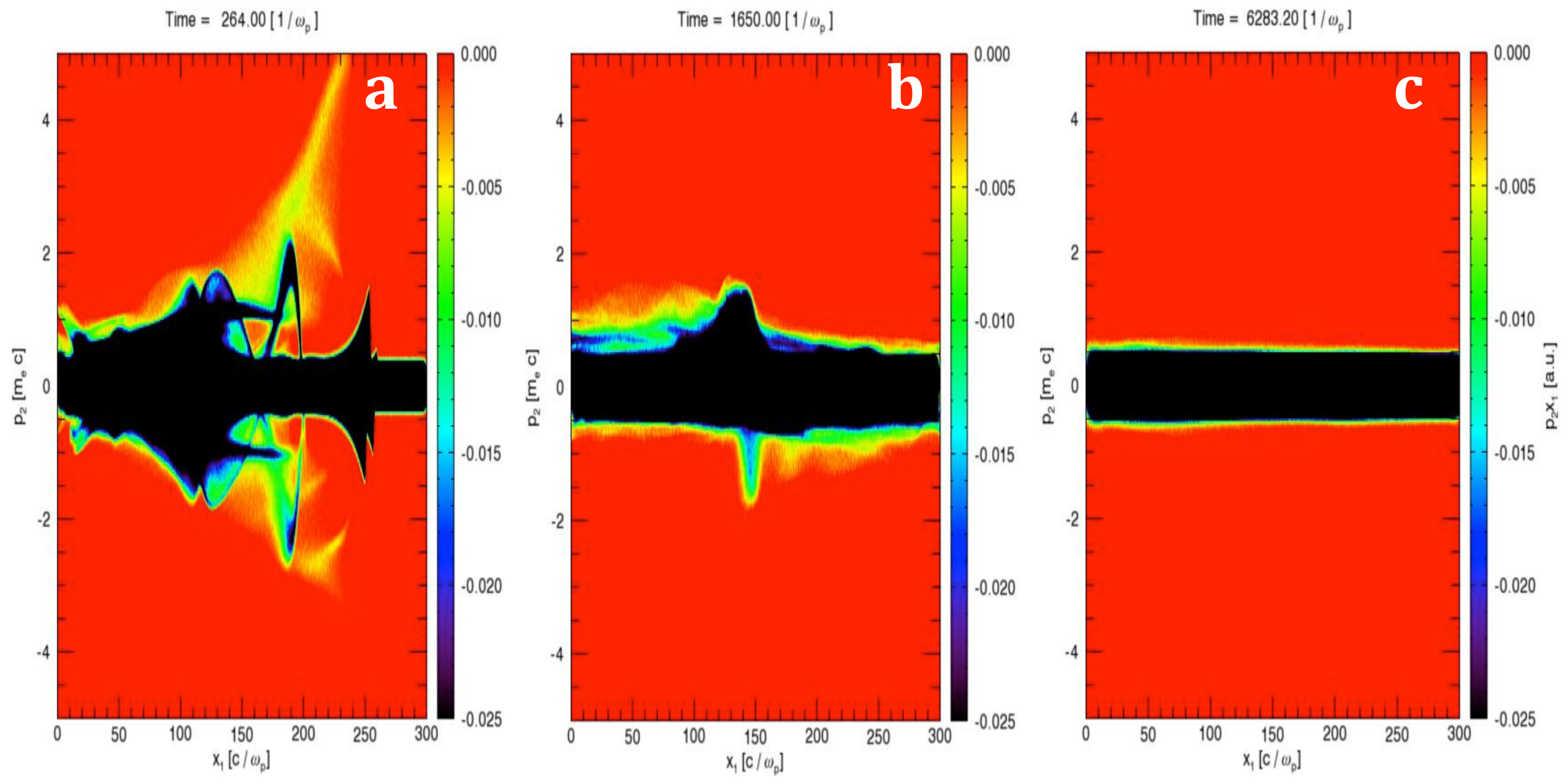}
	\caption{LTE snapshots of {\bf electron transverse momentum phase space} along the longitudinal direction at times, t= {\bf (a)} 264$\frac{1}{\omega_{pe}}$ {\bf (b)}  1650$\frac{1}{\omega_{pe}}$ {\bf (c)} 6283.20$\frac{1}{\omega_{pe}}$.}
	\label{LTE-transverse-momentum-evolution}
\end{figure}

At later times we also observe that neighboring plasma bubbles merge into each other and the electrons in the sheath execute motion around the merged ion cavity. The merged bubbles can be seen around $100\frac{c}{\omega_{pe}}$. The ion density at the same time as Fig.~\ref{LTE-electron-density-evolution}[b] is shown in Fig.~\ref{LTE-ion-density-evolution}[b]. From the ion density snapshot in Fig.~\ref{LTE-ion-density-evolution}[b] a feature can be observed in the background ions around $100\frac{c}{\omega_{pe}}$ which has shape similar to the merged bubble in electron density. The ion motion occurs as expected at the ion acoustic timescales as the Fig.~\ref{LTE-ion-density-evolution}[b] is shown is around $10\frac{1}{\omega_{ia}}$. After the onset of ion motion we observe structures with very low phase velocities in comparison to the plasma wave, as seen in Fig.~\ref{LTE-electron-density-evolution}[c] and Fig.~\ref{LTE-ion-density-evolution}[c]. In Fig.~\ref{LTE-ion-density-evolution}[c] we see ion expansion away from the axis and an ion channel is seen. 

\begin{figure}
	\centering
   	\includegraphics[width=\columnwidth]{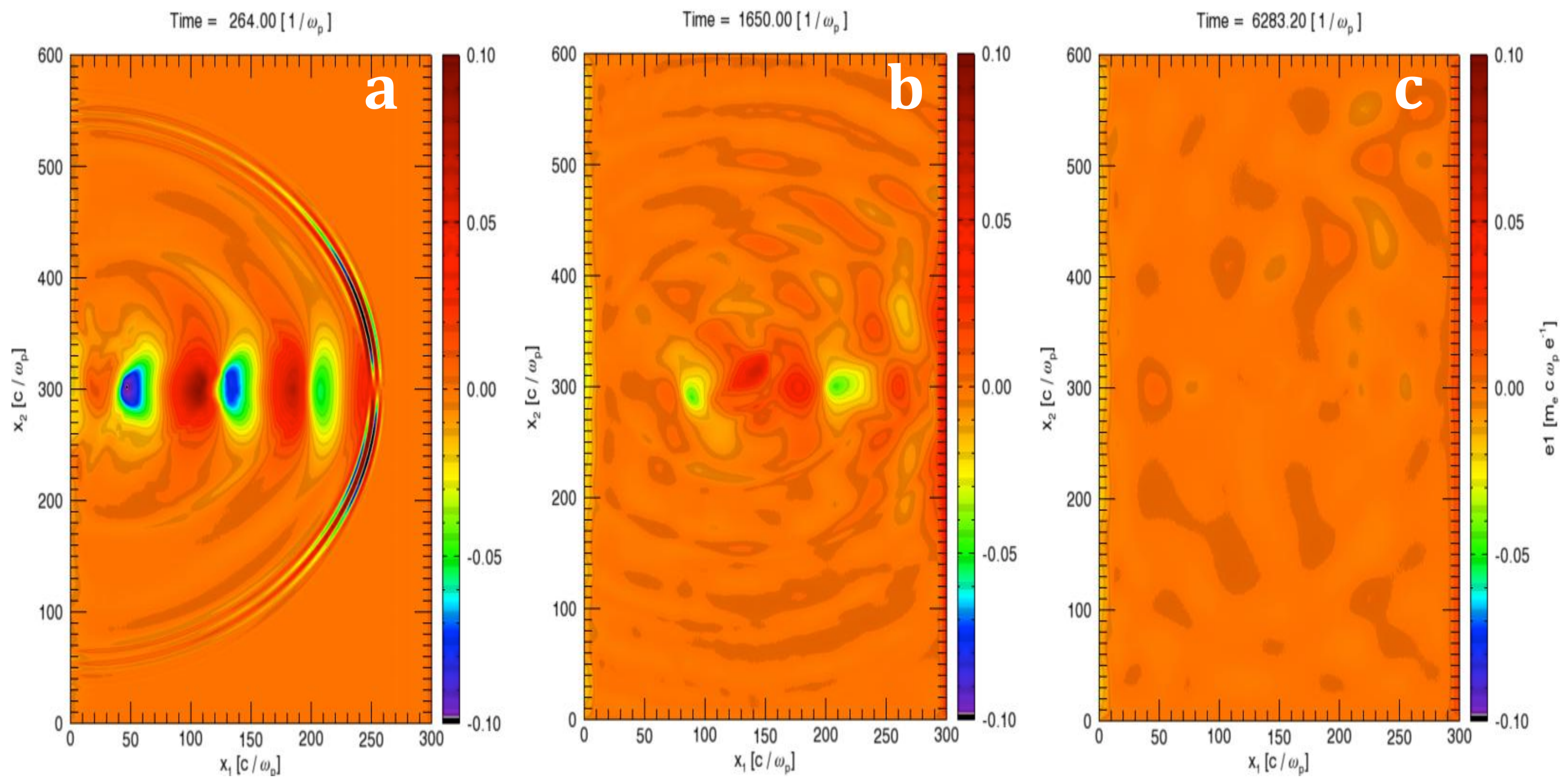}
	\caption{LTE snapshots of {\bf longitudinal electric field in real space} along the longitudinal direction at times, t= {\bf (a)} 264$\frac{1}{\omega_{pe}}$ {\bf (b)} 1650$\frac{1}{\omega_{pe}}$ {\bf (c)} 6283.20$\frac{1}{\omega_{pe}}$.}
	\label{LTE-longitudinal-field-evolution}
\end{figure}

\begin{figure}	
	\centering
   	\includegraphics[width=\columnwidth]{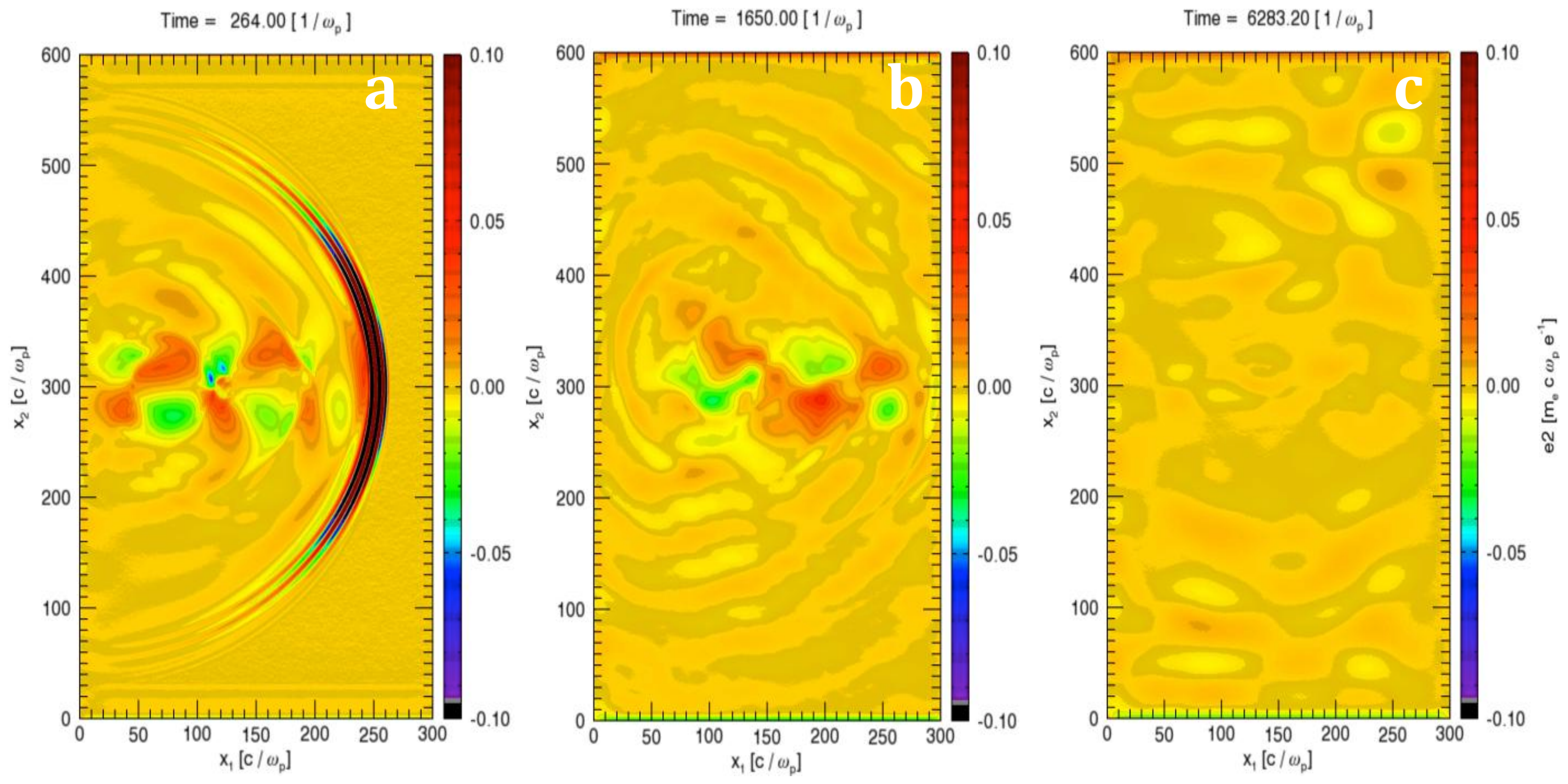}
	\caption{LTE snapshots of {\bf transverse electric field in real space} along the longitudinal direction at times, t = {\bf (a)} 264$\frac{1}{\omega_{pe}}$ {\bf (b)} 1650$\frac{1}{\omega_{pe}}$ {\bf (c)} 6283.20$\frac{1}{\omega_{pe}}$.}
	\label{LTE-transverse-field-evolution}
\end{figure}

In Fig.~\ref{LTE-longitudinal-momentum-evolution} and Fig.~\ref{LTE-transverse-momentum-evolution}, we show the momentum phase spaces corresponding to the density snapshots. In Fig.~\ref{LTE-longitudinal-field-evolution} and Fig.~\ref{LTE-transverse-field-evolution}, we show the fields in the real space corresponding to the density snapshots. It can be seen from the fields much after the short laser pulse ($10fs$ long) has passed by and the plasma wave has evolved and decayed there is still significant trapped electric field (Fig.~\ref{LTE-transverse-field-evolution}[b] and Fig.~\ref{LTE-longitudinal-field-evolution}[b]) in the merged bubble seen in the Fig.\ref{LTE-electron-density-evolution}[b]. This is the characteristic of a {\it caviton}. In the laser driven wake we can observe trapped laser radiation, however in beam driven case there is no driving radiation and hence the characteristic of beam-driven wake LTE differ from the laser-driven wake LTE.

\newpage

%\begin{align}
%\nonumber \epsilon_0 &= \frac{\Delta\omega_0}{\omega_0} \\
%\newline \epsilon(x,t) &= \epsilon_0\left(\frac{ct-x}{\theta}\right)H(ct-x)
%\label{laser-chirp}
%\end{align}
%\noindent The Heaviside step function $H(ct-x)$ ensures that the frequency chirping effect is observed at a point x in space only after the head of the pulse has reached that point, x (when $(ct-x)>0$).
%

\end{document}